\documentstyle[epsfig,longtable]{aipproc}

\begin{document}
\title{Ground-Based Gamma-Ray Astronomy}

\author{Michael Catanese}
\address{Iowa State University\thanks{Current address: Harvard-Smithsonian 
Center for Astrophysics, F.L. Whipple Observatory, P.O. Box 97, Amado, AZ 
85645.  This research is supported by the U.S. Department of Energy and 
by NASA}, Dept. of Physics and Astronomy, Ames, IA 50011}

\maketitle

\begin{abstract}
Ground-based $\gamma$-ray astronomy has become an active astrophysical
discipline with four confirmed sources of TeV $\gamma$-rays, two
plerionic supernova remnants (SNRs) and two BL Lac objects (BL Lacs).
An additional nine objects (one plerion, three shell-type SNRs, one
X-ray binary, and four BL Lacs) have been detected but have not been
confirmed by independent detections.  None of the galactic sources
require the presence of hadronic cosmic rays, so definitive evidence
of their origin remains elusive.  Mrk\,421 and Mrk\,501 are weak EGRET
sources but they exhibit extremely variable TeV emission with spectra
that extend beyond 10\,TeV.  They also exhibit correlations with lower
energy photons during multi-wavelength campaigns, providing tests of
emission models.  Next generation telescopes like VERITAS hold the
promise of moving this field dramatically forward.
\end{abstract}

\section*{Introduction}

Since the launch of the {\it Compton Gamma-Ray Observatory} ({\it
CGRO}), ground-based $\gamma$-ray telescopes have come to play an
important role in our understanding of the $\gamma$-ray sky.  In many
cases, it has required the results from both the ground and space to
properly interpret the observations of a particular source.  In this
context, I review the status of ground-based $\gamma$-ray astronomy
and consider the implications of these observations.  I will
concentrate on the results obtained with imaging atmospheric Cherenkov
telescopes because they have produced most of the scientific results
to date and because several papers in this proceedings address other
ground-based telescope results.  The interested reader is encouraged
to seek out more complete reviews that have recently been published
\cite{Catanese99,Ong99} for more information.  To save space, I will
not cite the original detection references for those objects that are
included in the review articles.

\section*{Galactic Sources}

Seven sources of very high energy (VHE, $E>250$\,GeV) $\gamma$-ray
emission associated with galactic objects have been detected at this
time: three plerionic supernova remnants (SNRs) (Crab Nebula,
PSR\,1706-44, and Vela), three shell-type SNRs (SN\,1006,
RXJ\,1713.7-3946 \cite{Muraishi99}, and Cassiopeia\,A
\cite{Puehlhofer99}), and the X-ray binary Centaurus\,X-3
\cite{Chadwick98}.  A summary of the VHE properties of these objects
is given in Table\,\ref{gal_srcs}.  The Crab Nebula and PSR\,1706-44
have been confirmed as sources of VHE $\gamma$-rays by detections from
independent groups.  The Crab Nebula has the highest VHE $\gamma$-ray
flux of these objects and this, along with its steady flux, has
established it as the standard candle of ground-based $\gamma$-ray
astronomy.  Because of this and because ground-based $\gamma$-ray
telescopes have a range of energy thresholds, I list source fluxes in
units of the Crab flux to make comparisons of source strength easier.

\begin{table}
\caption{Galactic Sources of VHE $\gamma$-rays.}
\label{gal_srcs}
\begin{tabular}{rccc}
\multicolumn{1}{c}{Source} & \multicolumn{1}{c}{Energy} & 
\multicolumn{1}{c}{Flux} & \multicolumn{1}{c}{Significance} \\
\tableline
Crab Nebula & $>$0.3\,TeV & 
 \multicolumn{1}{c}{1.26$\times$10$^{-10}$cm$^{-2}$s$^{-1}$} & 
 Conf.\tablenote{Significances are listed only for unconfirmed sources.} \\
PSR\,1706-44 & $>$1.0\,TeV & 0.38\,Crab & Conf. \\
Vela & $>$2.5\,TeV & 0.54\,Crab & 5.8\,$\sigma$ \\
SN\,1006 & $>$1.7\,TeV & 0.48\,Crab & 8.0\,$\sigma$ \\
RXJ 1713.7-3946 & $>$2.0\,TeV & 0.40\,Crab & 5.0$\sigma$ \\
Cassiopeia\,A & $>$0.5\,TeV & ??? & 4.7\,$\sigma$ \\
Centaurus\,X-3 & $>$0.4\,TeV & 0.24\,Crab & 6.5\,$\sigma$ \\
\end{tabular}
\end{table}

All three of the detected plerions are EGRET sources \cite{Hartman99},
but the GeV emission is predominantly or entirely pulsed, while the
TeV emission shows no evidence of pulsations.  This is consistent with
the VHE emission arising in the synchrotron nebulae of these
objects. Several groups have measured accurately the spectrum of the
Crab Nebula over an energy range spanning 250\,GeV to 50\,TeV. The
spectrum is fit well by a simple power law with differential spectral
index 2.5.  The sub-GeV flux measurements, combined with the VHE
measurements, are consistent with synchrotron self-Compton emission
models for a magnetic field of $\sim 160\,\mu$G \cite{Hillas98}.  An
interesting feature of the TeV emission detected from Vela is that the
peak in the TeV emission is located $\sim 0.14^\circ$ away from the
pulsar position, coincident with the birthplace of the pulsar.  An
upper limit of 0.40\,Crab above 300\,GeV for Vela \cite{Chadwick99a}
implies that its spectrum must be harder than $E^{-2.3}$.

\begin{figure}[t!]
\centerline{\epsfig{file=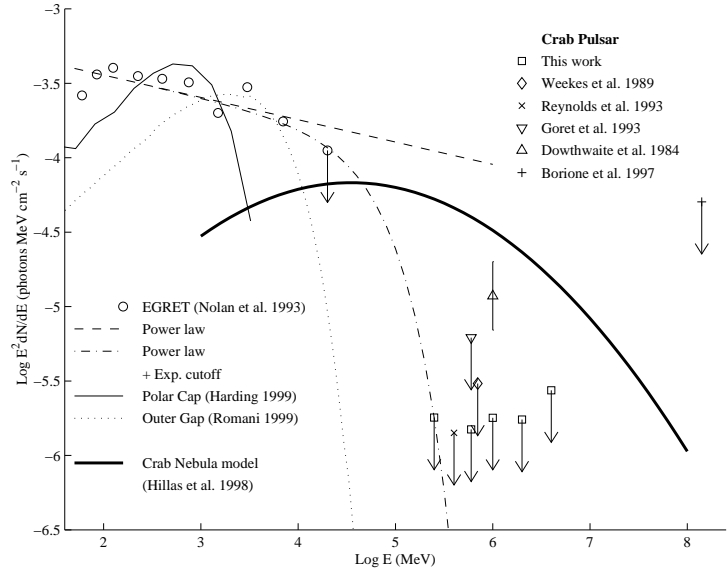,height=3.0in}}
\vspace{10pt}
\caption{The pulsed $\gamma$-ray spectrum of the Crab Nebula.  The
open circles are EGRET flux points and the points with arrows are
upper limits.  The thick solid line is a model of the unpulsed inverse
Compton spectrum.  The thin solid and dotted lines are model fits and
the dot-dashed line is an extension of the EGRET spectrum with a
60\,GeV exponential cut-off.  Figure from \protect\cite{Lessard99}.}
\label{crab_pulse}
\end{figure}

In addition to the plerions listed above, the pulsars
detected by EGRET have been searched for VHE emission.  None
have been detected, and the pulsed flux from these objects must have a
rapid decrease in power output between $\sim 1$\,GeV and 300\,GeV.  
Evidence for such cut-offs is
seen in the EGRET data for some of these objects \cite{Thompson97}, 
but for PSR\,1951+32, the power output increases up to at least 10\,GeV.  
The
Whipple collaboration's upper limit on this object is $\sim$0.02\,Crab
\cite{Hall99}, implying an extremely rapid fall off in the flux.
Similarly, recent upper limits derived for pulsed emission from the
Crab Nebula imply that if the cut-off is exponential, it must begin
below 60\,GeV, though this does not constrain the emission models (see
Figure~\ref{crab_pulse}) \cite{Lessard99}.

Shell-type SNRs are believed to be sources of $\gamma$-rays produced
by cosmic rays accelerated in the supernova shocks.  In support of
this, several EGRET sources are associated with shell-type SNRs
\cite{Esposito96}.  However, the EGRET detections alone are not enough
to claim definitively that the long-sought origins of cosmic rays have
been found.  Indeed, observations by the Whipple collaboration of
several of the shell-type SNRs associated with the EGRET sources
revealed no evidence of VHE $\gamma$-ray emission \cite{Buckley98}.
The limits derived from those observations imply that if the emission
seen by EGRET is from the SNR shells and is produced by the
interactions of cosmic rays, then the source cosmic-ray spectrum must
be steeper than the E$^{-2.3}$ or that the spectrum cuts off below
10\,TeV.

\begin{figure}
\centerline{\epsfig{file=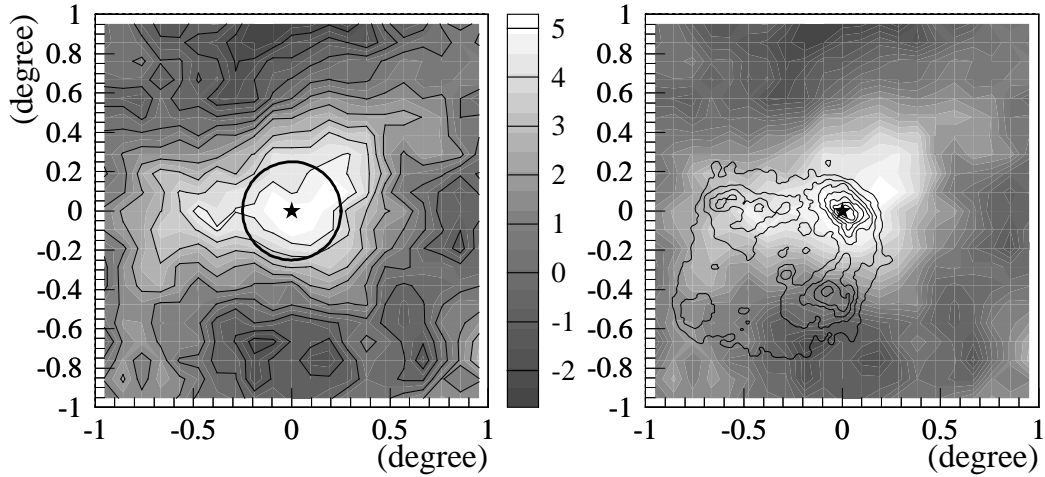,height=2.5in}}
\vspace{10pt}
\caption{CANGAROO observations of RXJ 1713.7-3946.  The left plot
shows the excess in TeV $\gamma$-rays and the right plot superimposes
the X-ray map of synchrotron emission over the TeV $\gamma$-ray map.
Figure from \protect\cite{Muraishi99}.}
\label{snr_fig}
\end{figure}

The three detected shell-type SNRs also do not require the presence of
hadronic cosmic rays to produce the $\gamma$-rays.  In all three
objects, X-ray synchrotron emission has been detected, implying the
presence of $> 10$\,TeV electrons.  Thus, the TeV detections can be
explained as inverse Compton emission.  This is supported by the
EGRET's non-detection of these objects and by the positional
coincidence of the TeV and X-ray synchrotron emission peaks in
SN\,1006 and RXJ\,1713.7-3946 (see Figure~\ref{snr_fig},
\cite{Muraishi99,Tanimori98}).  If the TeV emission from SN\,1006 is
produced from inverse Compton emission, the magnetic field in the
shock region must be $6.5 \pm 2 \mu$G \cite{Tanimori98}.

\section*{Extragalactic Sources}

Six BL Lacertae objects (BL Lacs) have been detected as sources of VHE
$\gamma$-rays: Markarian\,421 (Mrk\,421), Mrk\,501, 1ES\,2344+514,
PKS\,2155-304, 1ES\,1959+650 \cite{Nishiyama99}, and 3C\,66A.  A
summary of their properties is given in Table~\ref{agn_tab}.  The
results quoted in the table are values from the discovery papers.
Only Mrk\,421 and Mrk\,501 have been confirmed as VHE sources.  The
other objects have been detected with high significance in limited
time intervals, making confirmation difficult.  Mrk\,421,
PKS\,2155-304, and 3C\,66A are sources in the third EGRET catalog
\cite{Hartman99}.  Mrk\,501 was first detected on the ground but it
has recently been claimed as an EGRET source \cite{Kataoka99}.

\begin{table}
\caption{Extragalactic Sources of VHE $\gamma$-rays.}
\label{agn_tab}
\begin{tabular}{rcccc}
\multicolumn{1}{c}{Source} & $z$ & Energy & 
 Flux & \multicolumn{1}{c}{Significance} \\
\tableline
Mrk\,421 & 0.031 & $>$0.50\,TeV & 0.3\,Crab & Conf.\tablenote{Significances are
listed only for unconfirmed sources.} \\
Mrk\,501 & 0.034 & $>$0.30\,TeV & 0.08\,Crab & Conf. \\
1ES\,2344+514 & 0.044 & $>$0.35\,TeV & 0.11\,Crab & 5.2\,$\sigma$ \\
PKS\,2155-304 & 0.116 & $>$0.30\,TeV & 0.48\,Crab & 6.8\,$\sigma$ \\
1ES\,1959+650 & 0.048 & $>$0.90\,TeV & ??? & 3.7\,$\sigma$ \\
3C\,66A & 0.444 & $>$0.90\,TeV & 1.2\,Crab & 5.0\,$\sigma$ \\
\end{tabular}
\end{table}

\begin{figure}[t!]
\centerline{\epsfig{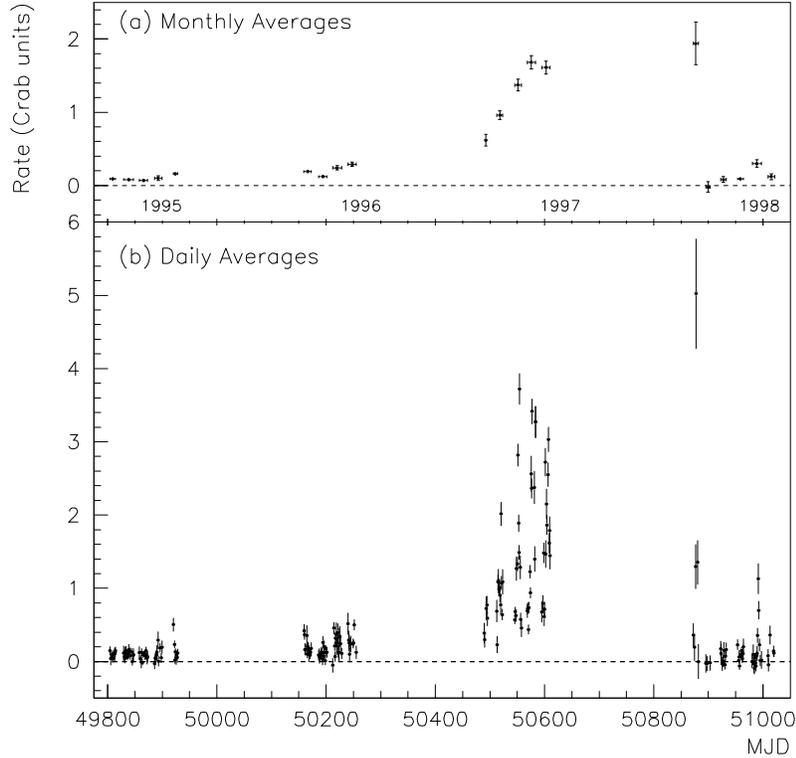}}
\vspace{10pt}
\caption{Whipple Observations of Mrk\,501 between 1995 and 1998.  The
top plot shows monthly average fluxes and the bottom plot shows
nightly average fluxes.  Figure from \protect\cite{Quinn99}.}
\label{m5var}
\end{figure}

The most distinctive feature of the VHE emission from Mrk\,421 and
Mrk\,501 is large amplitude, rapid variability.  For Mrk\,421, the
average flux does not change much from year to year.  Instead, flares
develop and decay on day-scales or less and drop to a baseline
emission level (if one exists at all) that is below the sensitivity of
current telescopes \cite{Buckley96}.  Fluxes from 0.1 to 10 times the
Crab flux have been detected and flares lasting as little as 30
minutes have been measured \cite{Gaidos96}.  For Mrk\,501, the flaring
appears to be somewhat slower and of lower amplitude than that seen in
Mrk\,421 \cite{Quinn99}.  The most prominent features of the
variability in Mrk\,501 are large changes in its average flux and
flaring activity, as shown in Figure~\ref{m5var}.  The yearly average
flux has varied from 0.08\,Crab in 1995 to 1.4\,Crab in 1997 and the
amount of day-scale flaring increases with increasing flux
\cite{Quinn99}.

Perhaps the most important development in the TeV results on Mrk\,421
and Mrk\,501 since the 4th Compton Symposium has been accurate
measurements of their spectra.  Observations of several high (1 --
10\,Crab) flux states between 1995 and 1996 from Mrk\,421 by the
Whipple collaboration show that the spectra are all consistent with a
simple power law with photon index $-2.54 \pm 0.03_{\rm stat} \pm
0.10_{\rm sys}$ over the energy range from 0.25 -- 10\,TeV
\cite{Krennrich99}.  Observations by the HEGRA collaboration of
Mrk\,421 in a lower ($<$1\,Crab) flux state in 1998 also indicate a
power law spectrum, but the spectral index is $-3.09 \pm 0.07_{\rm
stat} \pm 0.10_{\rm sys}$ over the energy range 0.5 -- 7\,TeV
\cite{Aharonian99a}.  This difference could reflect a change in
spectral index with flux, but neither Whipple nor HEGRA see evidence
of spectral variability within their respective data sets.  Further
study may help resolve these differences.

Unlike Mrk\,421, the spectrum of Mrk\,501 during its high state in
1997 is not consistent with a simple power law.  The Whipple
\cite{Krennrich99} and HEGRA \cite{Aharonian99b} collaborations derive
spectra of the form:
\begin{eqnarray*}
{{dN}\over{dE}} \propto & 
  E^{-2.22\pm 0.04_{\rm stat} \pm 0.05_{\rm sys} - 
  (0.47 \pm 0.07_{\rm stat})\log_{10}(E)} & {\rm (Whipple)} \\
{{dN}\over{dE}} \propto & 
 E^{-1.92\pm 0.03_{\rm stat} \pm 0.20_{\rm sys}} \times
 e^{-E/6.2 \pm 0.4_{\rm stat}(^{+2.9}_{-1.5})_{\rm sys}} & {\rm (HEGRA)}.
\end{eqnarray*}
The form of the curvature term in these two expressions reflects the
preferences of the authors, since the data from both groups are
indistinguishable when overlaid.  The average spectrum for Mrk\,501
measured by the CAT group is consistent with that observed by HEGRA
and Whipple \cite{Djannati99}, but the CAT data show evidence of
spectral hardening during high flux states while the Whipple and HEGRA
data do not.  Again, further study may resolve these issues.

\begin{figure}
\centerline{\epsfig{file=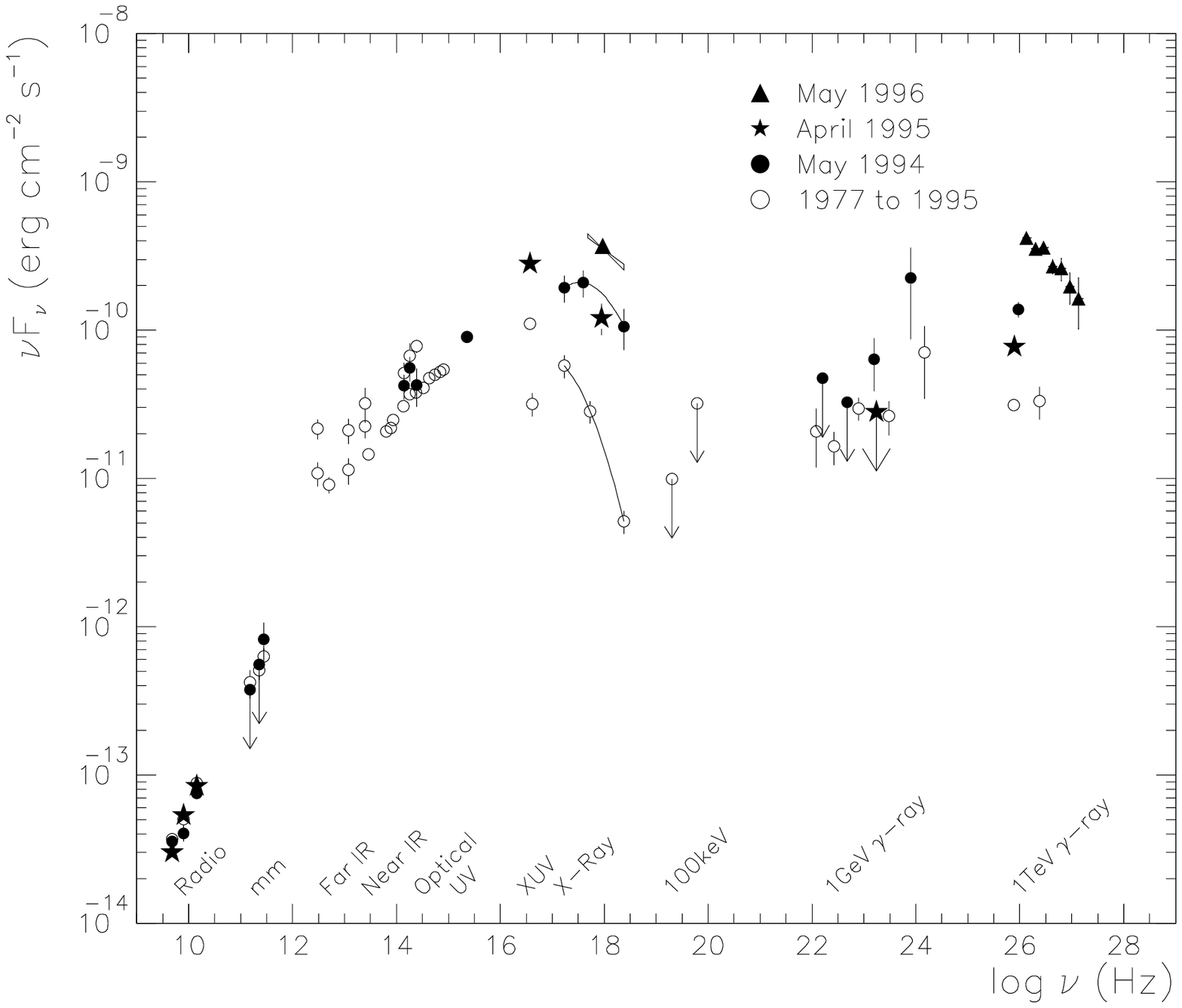,height=2.5in}\hspace*{0.1in}
\epsfig{file=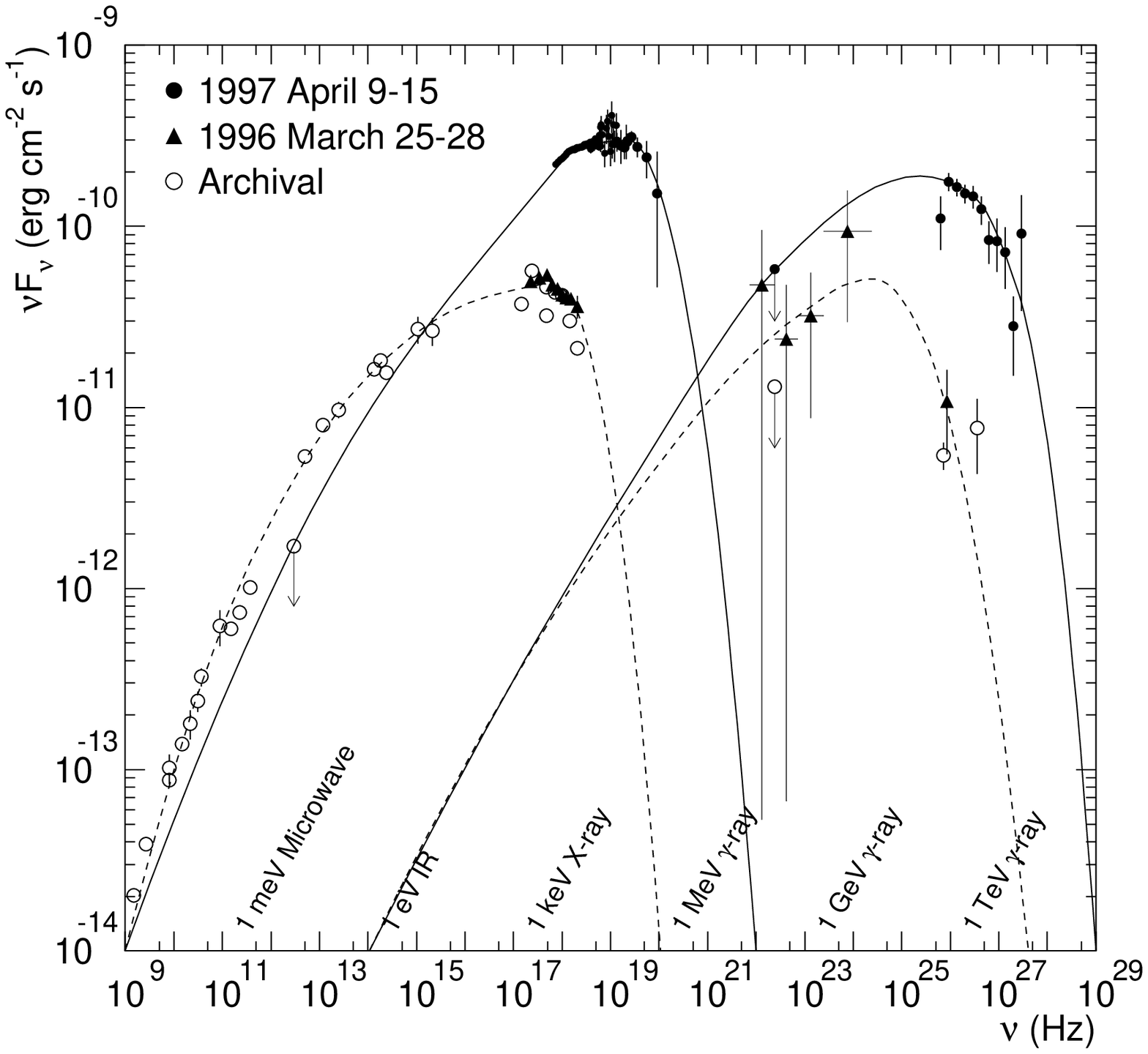,height=2.5in}}
\vspace{10pt}
\caption{Spectral energy distributions of Mrk\,421 and Mrk\,501 from
multi-wavelength campaigns and archival data.  Figure from
\protect\cite{Catanese99}.}
\label{m4_m5_sed}
\end{figure}

Mrk\,421 and Mrk\,501 have been the target of several intensive
multi-wavelength campaigns.  Observations of Mrk\,421 in 1995
\cite{Buckley96} and Mrk\,501 in 1997 \cite{Catanese97} revealed
day-scale correlations between the TeV $\gamma$-ray and X-ray
emissions, suggesting that both sets of photons derive from the same
population of particles.  The variability of the synchrotron emission
increases with increasing energy, and EGRET's lack of detected
variability in these studies suggests similar behavior for the high
energy emission. Thus, these flares seem to be caused primarily by
impulsive increases in the efficiency for acceleration of the highest
energy electrons.  This is not to say that the multi-wavelength
behavior of Mrk\,421 and Mrk\,501 is identical.  For example, in
Mrk\,421, the variability amplitude of the TeV $\gamma$-rays and
X-rays is comparable while for Mrk\,501, the variability amplitude is
larger in the TeV $\gamma$-rays.  More spectacular, is the difference
in the spectral energy distributions of these two objects, as shown in
Figure~\ref{m4_m5_sed}.  The spectrum of Mrk\,421 is typical of
high-frequency peaked BL Lacs: a synchrotron peak at $\sim$1\,keV
followed by a rapid drop-off.  Mrk\,501, on the other hand, appears to
be an extreme version of a high-frequency peaked BL Lac, as its
synchrotron spectrum peaked at 100\,keV in 1997, the highest ever
observed in a blazar.  Also, the power output at X-ray and TeV
energies in Mrk\,421 is approximately equal but for Mrk\,501, the TeV
power can be much less than in X-rays.

\begin{figure}[t!]
\centerline{\epsfig{file=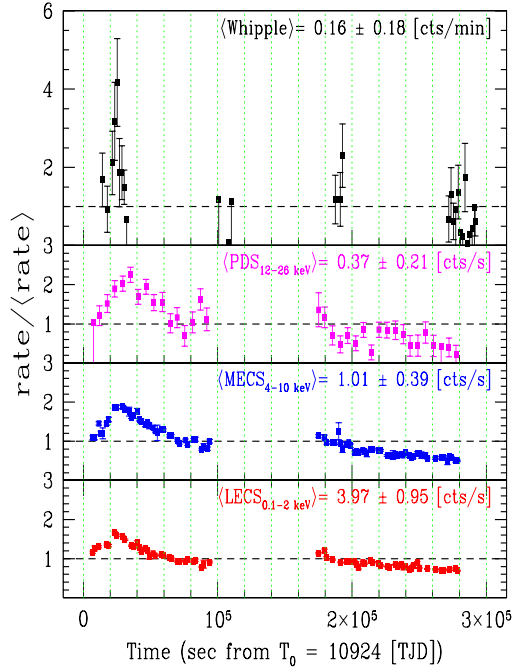,height=3.5in}}
\vspace{10pt}
\caption{The light curve of Mrk\,421 from 1998 April 21
to 24.  The data are normalized to their mean during the
observations (shown in each panel).  The errors listed indicate the
standard deviation of the data.  The Whipple data are for $E>2$\,TeV.
Figure from \protect\cite{Maraschi99}.}
\label{m4_1998}
\end{figure}

Those observations do not resolve any of the hour-scale flares known
to occur in Mrk\,421 and Mrk\,501.  In 1998, a campaign involving the
Whipple telescope and {\it BeppoSAX} had overlapping observations of
an hour-scale flare from Mrk\,421 as shown in Figure~\ref{m4_1998}
\cite{Maraschi99}.  The different energy bands exhibit a similar rise
time, but the TeV $\gamma$-ray flux appears to fall-off much faster
than the X-rays.  Thus, at the same time that the first hour-scale
correlations are seen in a TeV blazar, there is also evidence that the
TeV $\gamma$-rays and X-rays in Mrk\,421 may not be completely
correlated on all time-scales.

\section*{Conclusions}

From the previous paragraphs, it should be clear that ground-based
$\gamma$-ray astronomy has become a vibrant branch of
astrophysics.  There are established sources with well-measured
spectra and, in the case of the BL Lacs, variability light-curves.
There are several unconfirmed sources which lead me to believe that
more sources are to be found in this waveband.  And, there are some
controversies which need resolving (e.g., do the spectra in Mrk\,421 and
Mrk\,501 vary or not?) - which I interpret as a healthy sign of a
growing field.  

However, it is also clear that many questions remain unanswered.  For
example, none of the sources show conclusive evidence of cosmic-ray
acceleration.  Also, we do not know the particle content in blazar
jets, nor do we know where the emission spectra of most of the
EGRET-detected blazars cut-off.  Ground-based efforts, such as VERITAS
\cite{Weekes99}, will dramatically improve the measurements in the VHE
band.  Combined with the next generation of space-based $\gamma$-ray
telescopes (e.g., GLAST) and X-ray telescopes like Chandra and
Astro-E, many of these questions will hopefully be answered.

\end{document}